\documentclass[conference]{IEEEtran}
\IEEEoverridecommandlockouts
\usepackage{cite}
\usepackage{amsmath,amssymb,amsfonts}
\usepackage{algorithmic}
\usepackage{graphicx}
\usepackage{textcomp}
\usepackage{xcolor}
\def\BibTeX{{\rm B\kern-.05em{\sc i\kern-.025em b}\kern-.08em
    T\kern-.1667em\lower.7ex\hbox{E}\kern-.125emX}}
\begin{document}

\title{Image Scaling Attack Simulation: A Measure of Stealth and Detectability}

\author{\IEEEauthorblockN{Devon A. Kelly}
\IEEEauthorblockA{\textit{Department of Computer Science and Engineering} \\
\textit{Texas A\&M University}\\
College Station, TX, USA \\
devonkel4@tamu.edu}
\and
\IEEEauthorblockN{Sarah A. Flanery}
\IEEEauthorblockA{\textit{Department of Electrical and Computer Engineering} \\
\textit{Texas A\&M University}\\
College Station, TX, USA \\
sflanery@tamu.edu}
\and
\IEEEauthorblockN{Christiana Chamon}
\IEEEauthorblockA{\textit{Department of Electrical and Computer Engineering} \\
\textit{Virginia Tech}\\
Blacksburg, VA, USA \\
ccgarcia@vt.edu}}

\maketitle

\begin{abstract}
Cybersecurity practices require effort to be maintained, and one weakness is a lack of awareness regarding potential attacks not only in the usage of machine learning models, but also in their development process. Previous studies have determined that preprocessing attacks, such as image scaling attacks, have been difficult to detect by humans (through visual response) and computers (through entropic algorithms). However, these studies fail to address the real-world performance and detectability of these attacks. The purpose of this work is to analyze the relationship between awareness of image scaling attacks with respect to demographic background and experience. We conduct a survey where we gather the subjects’ demographics, analyze the subjects’ experience in cybersecurity, record their responses to a poorly-performing convolutional neural network model that has been unknowingly hindered by an image scaling attack of a used dataset, and document their reactions after it is revealed that the images used within the broken models have been attacked. We find in this study that the overall detection rate of the attack is low enough to be viable in a workplace or academic setting, and even after discovery, subjects cannot conclusively determine benign images from attacked images.
\end{abstract}


\section{Introduction}
Human cognition consists of obliviousness and a tendency to trust one another, rendering humans susceptible to deception, manipulation, and exploitation. Manipulation in today’s society most commonly shows itself in the digital world, as humans have not merely adopted social media as a tool for communication and connection, but an integral part of daily life  \cite{b1,b2,b3,b4,b5,b6,b7,b8,b9,b10,b11,b12,b13,b14,b15,b16,b17,b18,b19,b20,b21,b22,b23,b24,b25,b26,b27,b28}. In response to the COVID-19 pandemic, remote communication has become an essential part of everyday life, and social media platforms have increased in popularity. In the professional realm, remote work became normalized with virtual meetings, collaboration tools, and telecommuting technologies \cite{b8,b9,b10}. The routine exchange of images is a feature of contemporary communication, and as pictures are shared and received on a daily basis, their transmission also serves as a vector for social engineering attacks \cite{b1,b2,b3,b4,b5,b6,b7,b8,b9,b10,b11,b12,b13,b14,b15,b16,b17,b18,b19,b20,b21,b22,b23,b24,b25,b26,b27,b28}.

The crux of social engineering attacks is manipulation, a psychological vulnerability rather than an algorithmic flaw. Through deception, the victim feels rushed to make a decision from a seemingly reliable source, showing that even systems that contain layers of security can be infiltrated through social engineering \cite{b1,b2,b3,b4,b5,b6,b7,b8,b9,b10,b11,b12,b13,b14,b15,b16,b17,b18,b19,b20,b21,b22,b23,b24,b25,b26,b27,b28,b29,b30,b31,b32,b33,b34,b35,b36,b37,b38,b39,b40,b41,b42,b43,b44,b45}. Despite the frequency of security breaches, not all social engineering attacks are immediately identified, as social engineers disguise their ulterior motives behind phishing emails, phone calls, or even face-to-face encounters \cite{b15,b16}. As such, the nature of these attacks raises a demand for constant vigilance and increased layers of protection on cyber-physical systems, including photo-editing software.

In the context of machine learning (ML), image scaling takes advantage of the static methodology of which different ML libraries downscale images to fit the model, independent of the input image \cite{b30,b31,b37}. An application of this attack includes teaching a model to provide a fixed response when it sees a certain pattern, usually contrary to the original output the model would have provided. As a result, an attacker can learn the input size and downscaling algorithm to determine where in the image to inject pixels. For example, let’s consider a self-driving car being trained on various road signs, and an attacker injects a picture of a stop sign in a training image that contains a clear road, in order to trick the self-driving car into not stopping at stop signs. Several defense mechanisms have been implemented against such attacks, including the corresponding detection mechanisms that operate by measuring the variance of the pixels surrounding where an attack is expected to take place, and comparing the expected value with the calculated variance \cite{b48}. However, given that the attacker and victim of such attacks are human, it is unclear whether the ignorance and lack of awareness of humans are contributing factors to the success of such attacks. In the present paper, we conduct a survey where we analyze the subjects’ experience in cybersecurity, record their responses to compromised images as a result of image scaling attacks, and make note of their reactions after we reveal to them that these images have been attacked. The purpose of this survey is to provide a better understanding on how people will likely respond if their dataset has been hijacked through the use of image scaling attacks. This division of sections serves to simulate a realistic scenario where the computer scientist debugging a ResNet model may not be aware of a security breach or a tampered dataset.

The rest of this paper is organized as follows. Section 2 describes the methodology used to carry out the survey, Section 3 displays the results of the survey, and Section 4 concludes this paper with discussion on the survey results.

\section{Methodology}
The target audience of the present study is participants that have experience in the software development and data science fields. The inclusion criteria consists of faculty, staff, and students who are at least 18 years old and pursuing a degree in computer science or a field related to machine learning. We would like to note that no personally identifying information will be collected, i.e. the survey is completely anonymous. The survey is sent through emails to relevant departmental Listservs, and advertised in person within classes during the designated lecture and lab periods.

The subjects are asked to fill out a Qualtrics survey consisting of the following sections: Demographics and Background, Minimal Information Survey, Debrief, and follow-up survey.

The Demographics and Background section gathers the age, gender, race, and experience in computer science, machine learning, and cybersecurity whether from work, academia, military, law enforcement, and/or hobbyism. 

The Minimal Information Survey begins with a short explanation of the simulated scenario: the subject is attempting to figure out why a ResNet model is performing poorly on a cats and dogs dataset \cite{b49}. The survey inquires the subjects on what part of the model would be causing problems: the model itself, the hyperparameters used, the training images from the dataset, or no issue at all, and why the subject made their selection. To assist in the selection process, subjects are given the detailed model layout (Tensorflow's implementation of ResNet50, a GlobalAveragePooling2D layer, a 128 node dense layer, and a 1 node dense layer), knowledge that the starting weights are the same as the subject's simulated coworker, and a specific selection of the training images to supplement their decision-making process.

\begin{figure}
    \centering
    \includegraphics[width=\linewidth]{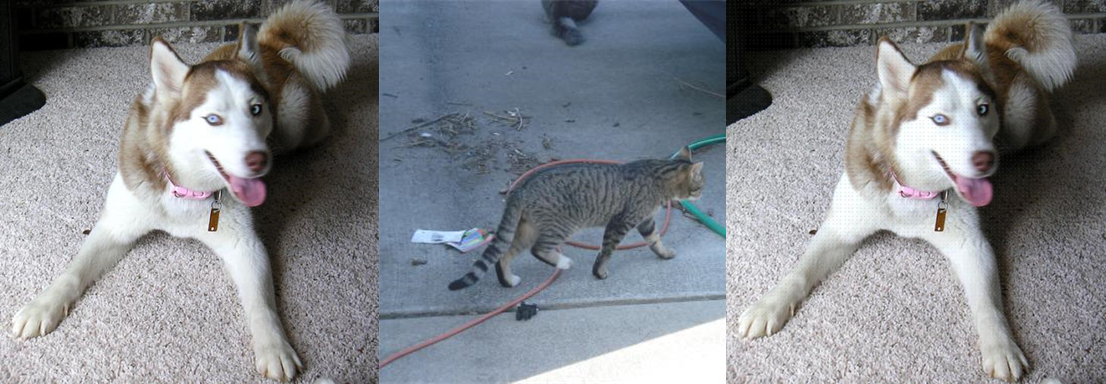}
    \caption{Example of an image infected by an image scaling attack. The original image (left) is a photograph of a dog. The middle image is injected into the attacked image (right). When scaled down, the attacked image becomes an image of a cat (middle).}
    \label{imagescalingdemo}
\end{figure}

We would like to note that in this phase, the subjects will not be told that several images have been compromised. An example is shown in Figure~\ref{imagescalingdemo}, where a cat photo has been injected into a dog photo, yet the resulting photo appears visually to be the dog, as it has not been scaled down. The original and target images were chosen such that the two contrast each other, creating a near-worst-case-scenario, where images containing contrasting features were chosen to be the original image and injecting image respectively.

The debrief section of the survey observes the awareness of the subject to image scaling attacks along with other attacks that target the preprocessing phase of machine learning models to manipulate results. In this phase, subjects are given the explanation that some of the images used in the inception of the model have been compromised by a third party, causing the problems mentioned in the Minimal Information phase--more specifically, subjects will be informed about image scaling attacks and how they operate, including the small artifacts within compromised images, such as the dots in Figure~\ref{imagescalingdemo}. Subjects are then further informed that some images have been affected by this attack, i.e. pictures of dogs appear to the model as a cat and vice versa, and that the reason for the poor performance of the model can be explained by the failed recognition of these images. The subjects will then be inquired on their prior awareness of image scaling attacks, and whether an attack was considered as a possibility during the Minimal Information phase.

\begin{figure}
    \centering
    \includegraphics[width=\linewidth]{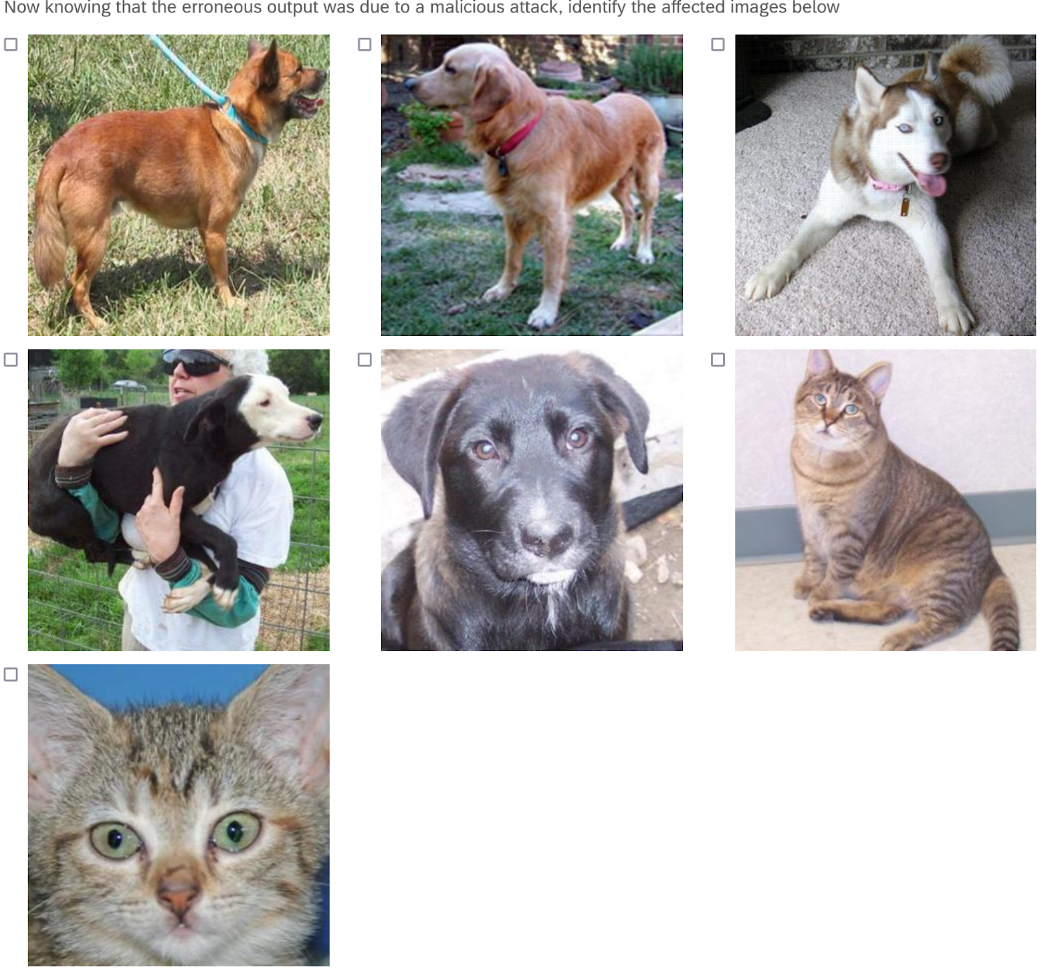}
    \caption{Selected images for the quiz. In this non-randomized order, the first 3 images have been injected via image scaling attacks. The rest of the 7 images are benign and unaffected. Users are tasked with selecting the attacked images and leaving the benign images unchecked.}
    \label{postquizoptions}
\end{figure}

After the debrief, subjects will take a multiple-choice quiz on more images from the dataset, consisting of three infected images and four benign images displayed in a randomized order. An example of this is shown in Figure~\ref{postquizoptions}. The result will determine whether they can identify attacked images after being aware of image scaling attacks and their presence, along with the details regarding ways to determine if an image has been infected. The post-debrief analyzes how subjects are able to determine the integrity of potentially-attacked images after gaining awareness.

By collecting responses from subjects with varying levels of academic and industry experience, we analyze the true detectability of image scaling attacks. We also analyze each groups’ ability to identify such attacks, thus indicating the overall population’s awareness of such attacks during the training step of machine learning.

\section{Results}
\begin{figure}
    \centering
    \includegraphics[width=\linewidth]{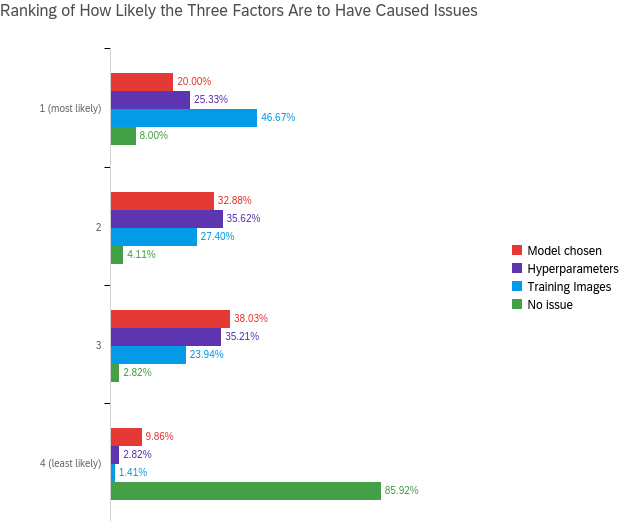}
    \caption{Responses from users by percent regarding the rankings of the potential problems within the neural network model presented during the survey.}
    \label{initialbars}
\end{figure}

We have gathered 129 responses, but only 78 have met the qualifying requirements in Section~\ref{meth}. Figure~\ref{initialbars} illustrates a quantitative comparison of what users believed to be the issue with the image in the Minimal Information survey (see Section 2). Table~\ref{initialtable} displays the statistics of the same data points, showing the mean ranking of each selection along with standard deviation. The subjects had an average industry experience of .38 years and 3.29 years of computer science experience. 92.13\% of respondents have some college, 7.09\% have their Bachelors degree, and 1 respondent had their PhD.

\begin{figure}
    \centering
    \includegraphics[width=\linewidth]{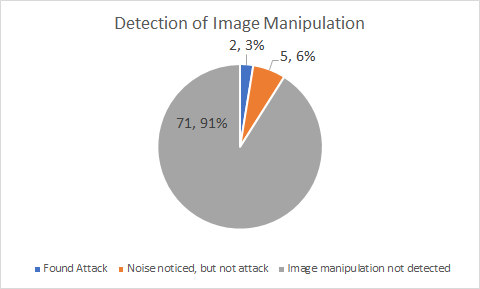}
    \caption{Percentage of users who detected problems with the dataset or immediately detected the attack. This data was taken from the minimal information survey (see Section~\ref{meth}).}
    \label{piechart}
\end{figure}

While the metrics in Figure~\ref{initialbars} and Table~\ref{initialtable} indicate that most subjects found an issue with the training images, but Figure~\ref{piechart} shows that only 2 directly pointed out that the images had been tampered with, and 5 noted that the noise in the image may have caused errors within the model. The remainder of the explanations for their selections included comments such as “the dogs had too many cat features”, “the cats had too many dog features”, “there was a sampling bias”, and other reasons that did not indicate a perception of an attack having been implemented. Less than 10\% of subjects noticed the manipulation done to the images with all photos shown to subjects being infected with an exaggerated attack with greater artifacts. Additionally, only 2 out of the 78 subjects noted that the attack was present, thus in the present experiment, the attack had a true discovery rate of 3\%.

\begin{table*}
\centering
\caption{Statistical information of the data in Figure~\ref{postquizoptions}. The black arrow designates an ordering of the mean ranking, where a lower mean indicates a higher ranking of likelihood in the responses.}
\begin{tabular}{|c|c|c|c|c|c|}
\hline
\textbf{Field} & \textbf{Minimum} & \textbf{Maximum} & \textbf{Mean} & \textbf{Std Deviation} & \textbf{Variance} \\ \hline
Training Images & 1.00 & 4.00 & 1.78 & 0.85 & 0.72 \\ \hline
Hyperparameters & 0.00 & 4.00 & 2.11 & 0.87 & 0.76 \\ \hline
Model chosen & 1.00 & 4.00 & 2.36 & 0.91 & 0.83 \\ \hline
No issue & 0.00 & 4.00 & 3.59 & 0.99 & 0.98 \\ \hline
\end{tabular}
\label{initialtable}
\end{table*}

\begin{figure}
    \centering
    \includegraphics[width=\linewidth]{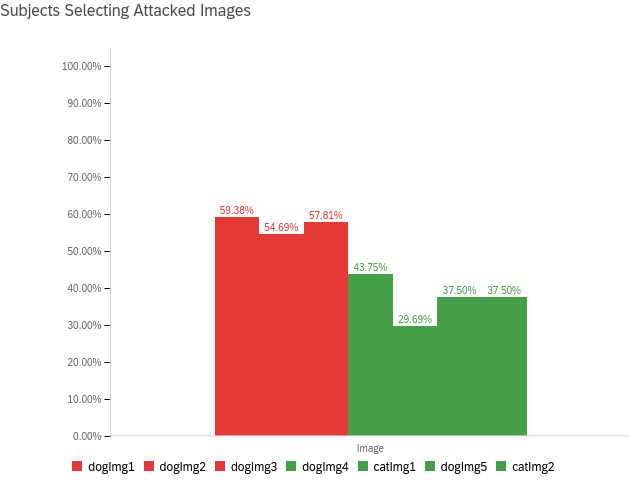}
    \caption{Percentage of respondents that have identified an attacked/benign image in Figure~\ref{postquizoptions}. The red bars represent the correct identification of the attacked images, and the green bars represent the correct identification of the benign images.}
    \label{quizbars}
\end{figure}

After explaining to the subjects what image scaling attacks are and how this form of attack had been applied to the dataset in the Debrief section (see Section 2), subjects were quizzed on which three of the seven images had been affected by image scaling. As shown in Figure~\ref{quizbars}, 57.29\% of subjects correctly identified the three attacked images. However, 37.11\% of subjects incorrectly marked the remaining 4 benign images as attacked. These metrics were calculated by averaging the results of the first three images and the last four, respectively, from Figure~\ref{quizbars}. This high amount of false positives and false negatives indicate that even after detection, many images still blend in, evading detection. Additionally, the false positive marking of benign images after detection could be caused by anticipation of the noise being present. In a scenario where the injecting images were chosen with similar color scheme to the original images, it would be likely that the numbers would stratify more with more false negatives and false positives, respectively. This is because subjects are able to determine the infected nature of the images through the artifacts left behind by the attack itself. However, if a more well-blended output image was created through smart pairing of original images and target images, then the artifact traces would be less clear to both the human eye and any algorithm attempting to detect the attack's presence.

\begin{figure*}
    \centering
    \includegraphics[width=\linewidth]{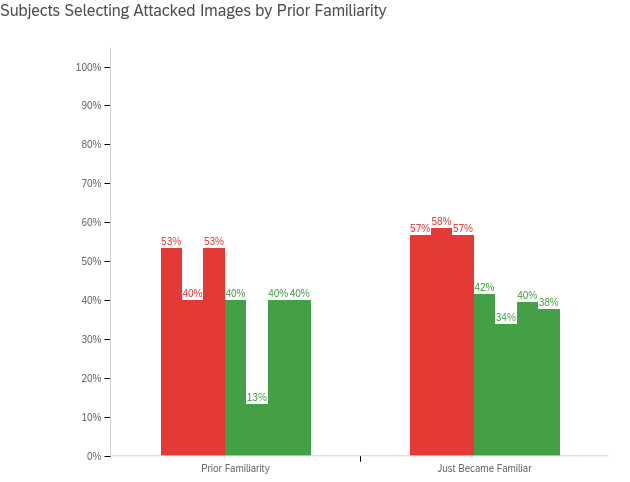}
    \caption{Comparison between respondents who have identified an attacked/benign image prior to the survey, with respect to prior familiarity. Similarly to Figure~\ref{quizbars}, the red bars represent the correct identification of the attacked images, and the green bars represent the correct identification of the benign images.}
    \label{quizbarsbyfamiliar}
\end{figure*}

When subjects were inquired about their thought processes during the first section, 48.72\% of subjects claimed to have considered an attack being present in the minimal information section, and only 21.79\% of subjects were familiar with Image Scaling (or similar preprocessing) attacks. Figure~\ref{quizbarsbyfamiliar} shows that subjects who had only just become aware of this type of attack had a higher rate of detecting the attacked images. As a trade-off, the subjects who just learned about the attack had a higher false positive rate than the subjects with prior knowledge of the attack.

\section*{Conclusion}
We conducted a survey to simulate the debugging experience of a neural network model while under an Image Scaling attack. We performed the survey by guiding subjects through a scenario which described the model's setup, parameters, and input. The input in this case had been modified through Image Scaling attacks, specifically choosing images that would display the worst case result of a random selection of images from opposing datasets. When subjects were prompted regarding which part of the setup was the problem, most subjects identified that the images were the source of the problem, yet only 9\% of them noticed the noise in the image from the image scaling attack, and only 2 directly stated that the images had been tampered with.

With the low detection rate of the attack and even lower detection of the noise artifacts in the image, this experiment suggests that the image scaling attack is effective in a workplace environment, especially with the low familiarity that the subjects had with the ability to attack the input of machine learning models.

We also showed that, after detection, subjects have difficulty differentiating attacked images from benign images, which rendered both high false positive and high false negative rates. To the extent that a little under half of the subjects failed to identify the attacked image, and a similar proportion of subjects marked the benign images as attacked. This indicates that the attack is not only effective from a detection standpoint, but that post-detection it is hard to determine the integrity of images.

If a potential attacker is to use an image scaling attack, whether through a malware that mixes dataset images together, by releasing poisoned datasets, or other means, then there would be significant damage to the workplace productivity for the models attempting to use the tampered images. It would also potentially cause many images to be falsely removed for being potentially tampered with post-discovery, or eliminating good datasets for being falsely assumed to be infected.

\end{document}